# A quantum liquid of magnetic octupoles on the pyrochlore lattice

Romain Sibille[1,*], Nicolas Gauthier[2], Elsa Lhotel[3], Victor Porée[1], Vladimir Pomjakushin[1], Russell A. Ewings[4],

Toby G. Perring[4], Jacques Ollivier[5], Andrew Wildes[5], Clemens Ritter[5], Thomas C. Hansen[5], David A. Keen[4],

Gøran J. Nilsen[4], Lukas Keller[1], Sylvain Petit[6,*] & Tom Fennell[1]

[1]Laboratory for Neutron Scattering and Imaging, Paul Scherrer Institut, 5232 Villigen PSI, Switzerland. [2]Stanford Institute for Materials and Energy Science, SLAC National Accelerator Laboratory and Stanford University, Menlo Park, California 94025, USA. [3]Institut Néel, CNRS–Université Joseph Fourier, 38042 Grenoble, France. [4]ISIS Pulsed Neutron and Muon Source, STFC Rutherford Appleton Laboratory, Harwell Campus, Didcot, OX11 0QX, UK. [5]Institut Laue-Langevin, 71 avenue des Martyrs, 38000 Grenoble, France. [6]LLB, CEA, CNRS, Université Paris-Saclay, CEA Saclay, 91191 Gif-sur-Yvette, France. *email: romain.sibille@psi.ch ; sylvain.petit@cea.fr

**Spin liquids are highly correlated yet disordered states formed by the entanglement of magnetic dipoles[1]. Theories typically define such states using gauge fields and deconfined quasiparticle excitations that emerge from a simple rule governing the local ground state of a frustrated magnet. For example, the '2-in-2-out' ice rule for dipole moments on a tetrahedron can lead to a quantum spin ice in rare-earth pyrochlores – a state described by a lattice gauge theory of quantum electrodynamics[2-4]. However, *f*-electron ions often carry multipole degrees of freedom of higher rank than dipoles, leading to intriguing behaviours and 'hidden' orders[5-6]. Here we show that the correlated ground state of a $Ce^{3+}$-based pyrochlore, $Ce_2Sn_2O_7$, is a quantum liquid of magnetic octupoles. Our neutron scattering results are consistent with the formation of a fluid-like state of matter, but the intensity distribution is weighted to larger scattering vectors, which indicates that the correlated degrees of freedom have a more complex magnetization density than that typical of magnetic dipoles in a spin liquid. The temperature evolution of the bulk properties in the correlated regime below 1 Kelvin is well reproduced using a model of dipole–octupole doublets on a pyrochlore lattice[7-8]. The nature and strength of the octupole–octupole couplings, together with the existence of a continuum of excitations attributed to spinons, provides further evidence for a quantum ice of octupoles governed by a '2-plus-2-minus' rule. Our work identifies $Ce_2Sn_2O_7$ as a unique example of a material where frustrated multipoles form a 'hidden' topological order, thus generalizing observations on quantum spin liquids to multipolar phases that can support novel types of emergent fields and excitations. The composite 'dipole–octupole' nature of the degrees of freedom and their evolution as a function of temperature or magnetic field provide new ways of controlling emergent phenomena in quantum materials.**

In materials with strong electronic correlations, numerous phenomena and associated phase transitions are explained in terms of ordered structures where individual degrees of freedom can be described using the first term of a multipolar expansion (dipole moments). However, further terms are required to explain an increasing number of novel phenomena in condensed matter physics. Such multipole moments, where order is 'hidden' from usual probes, characterize the anisotropic distributions of electric and magnetic charges around given points of the crystal structure. Multipoles can arise at the atomic scale from spin–orbit coupling, leading to well-characterized examples of 'hidden' orders such as in $Ce_{1-x}La_xB_6$ or $NpO_2$ (refs. 5,6). In the heavy-fermion material $URu_2Si_2$, multipoles are proposed to explain a mysterious electronic state appearing below a phase transition involving a large entropy change[5,6,9]. Multipoles can also emerge from atomic clusters, such as in the spin-liquid regime of $Gd_3Ga_5O_{12}$ with effective magnetic multipoles formed from 10-spin loops[10]. A number of studies have also pointed to the role of unconventional odd-parity multipoles to explain phase transitions that break both space inversion and time reversal. This was discussed for example in the context of the pseudogap region of high-transition-temperature superconductors[11-12] and in related iridates[13], as well as in metals with unconventional properties[14-15] and in magnetoelectric insulators[16-17].

Although sometimes hosting exotic behaviours, materials with multipolar orders can still be explained in the framework of symmetry-breaking phase transitions. Understanding physics beyond this paradigm is a major challenge, to which a significant part is contributed by spin liquids forming topological orders[18]. Such phases are defined by patterns of long-range quantum entanglements and fractionalized excitations – quasiparticles that cannot be constructed as combinations of the elementary constituents of the system. In 2- and 3-dimensional magnets, theoretical models of quantum spin liquids are now well-established and their applicability extends to a variety of frustrated spin systems[1]. However, connecting these predictions with experimental results remains difficult[19]. Measurements of fractionalized excitations were reported in model materials of Heisenberg spins with $S = 1/2$ or 1, such as the kagome lattice of $Cu^{2+}$ ions in the 2-dimensional herbertsmithite mineral[20] or the 3-dimensional pyrochlore lattice of $Ni^{2+}$ ions in $NaCaNi_2F_7$ (ref. 21). Alternatively, indications for quantum spin liquid physics exist in materials where the low-temperature degrees of freedom are $S_{eff} = 1/2$ pseudo-spins of a spin–orbital-entangled ground state doublet, such as in the 5$d$-electron $H_3LiIr_2O_6$ (ref. 22) and 4$f$-electron $YbMgGaO_4$ (ref. 23) materials. However, these materials can still be described on the basis of magnetic dipoles where the effect of spin–orbit coupling is limited to the creation of spin-space anisotropies.

In magnetic rare-earth pyrochlores, distinct types of spin–orbital-entangled doublets of $S_{eff}$ = 1/2 are allowed by the local trigonal $D_{3d}$ symmetry (ref. 24). Ions with odd numbers of 4$f$ electrons have their ground-state degeneracy imposed by Kramers' theorem, in which case the pseudo-spins have components that transform either like magnetic dipoles only or like magnetic dipoles and magnetic octupoles (rank-3 multipoles)[24]. Alternatively, pseudo-spins based on 'non-Kramers' doublets owe their degeneracy to the site symmetry and support magnetic dipole and electric quadrupole components[24]. Interactions between pseudo-spins sometimes involve only one component, in which case all terms in the Hamiltonian commute and the resulting phases are classical[1]. A famed example is the spin-ice state[25] found in $Ho_2Ti_2O_7$ and $Dy_2Ti_2O_7$, where magnetic dipoles of moment value $m$ ~ 10 $\mu_B$ correspond to the $z$ component $\tau_i^z$ of each pseudo-spin – the direction along the local trigonal axis connecting the centre of the elementary tetrahedron with its vertex. Interactions in this case are dominated by classical dipole–dipole forces acting as ferromagnetic first-neighbour couplings ($E_{dipolar}$ ~ +2 K), as required to make the '2-in-2-out' ice constraint effective. The result is a manifold of degenerate ice-rule states understood as the vacuum of an emergent Coulomb gauge field. However, in the absence of significant interactions $\mathcal{J}^{\pm}$ between pseudo-spin components $\tau_i^{\pm}$ transverse to $\tau_i^z$ (ref. 26), spin ices freeze and fall out of equilibrium at low temperature. Introducing zero-point fluctuations in order to stabilize a quantum spin ice requires a perturbative term[4], which, for instance, was proposed in non-Kramers $Pr^{3+}$-based pyrochlores where $\mathcal{J}^{\pm}$ corresponds to interactions between electric quadrupoles[27]. Here we reveal experimental signatures of a state where the dominant term in the Hamiltonian couples magnetic octupoles ($\tau_i^y$) while weaker interactions between magnetic dipoles ($\tau_i^z$) play the role of the transverse exchange, resulting in an octupolar quantum spin liquid predicted by theory[7-8]. This is realized in $Ce_2Sn_2O_7$, where muon-spin relaxation measurements have ruled out the presence of magnetic order at least down to a temperature $T$ of 0.02 K, while signatures of a correlated state were shown to appear in the bulk properties below $T$ ~ 1 K – that is 50 times higher, therefore pointing to a mysterious quantum phase[28].

We first report inelastic neutron scattering measurements at a low temperature $T$ of 6 K (Fig. 1**a-c**), probing the crystal-field states of $Ce^{3+}$ ions in our phase-pure sample of formula $Ce_2Sn_2O_{7.00\pm0.01}$ (see Extended data). Measurements with an incident energy $E_i$ of 150 meV reveal two magnetic excitations that are easily identified at energy transfers $E$ ~ 51 and 110 meV. These excitations are transitions from the ground to the excited Kramers doublets within the $J$ = 5/2 multiplet of $Ce^{3+}$, which itself results from the spin–orbital entanglement of the electron spin $S$ = 1/2 and orbital angular momentum $L$ = 3. Higher-multiplet transitions are also visible from the data (Fig.

1**b-c**). We have fitted the neutron data to a crystal-field Hamiltonian using the complete set of 14 intermediate-coupling basis states of the $J = 5/2$ and $J = 7/2$ multiplets (Fig. 1**d**), as required by the comparable strengths of the crystal-field and spin–orbit interactions in light rare-earths (see Extended data for full description of the fitting). The resulting crystal-field parameters allow to reproduce the high-temperature behaviour of the magnetic susceptibility $\chi$ to a high accuracy, as shown in Fig. 1**e**, where the effective magnetic dipole moment $\mu_{eff} \propto (\chi \times T)^{1/2}$ – a sharply-discriminating quantity – is plotted as a function of temperature. At temperatures much below the first excited crystal-field level but above the correlated regime, that is between 1 and 10 K, the magnetic dipole moment is $m \sim 1.2\ \mu_B$. This value translates into classical dipole–dipole couplings of about 0.025 K ($E_{dipolar} \propto m^2$), which is small compared to the scale of the dominant interactions and confirms that the low-temperature correlated state originates from quantum-mechanical exchange interactions[28]. The wavefunction of the ground-state Kramers doublet $|\pm\rangle$ obtained from the neutron data is essentially any linear combination of the $|m_{J_z} = \pm 3/2\rangle$ states, which corresponds to the case where pseudo-spins form a composite object, carrying both a magnetic dipole and a magnetic octupole (ref. 8).

We now focus on the nature of the correlated ground state using energy-integrated neutron scattering data recorded at temperatures inside and outside this regime, which by difference gives access to the Fourier transform of the total magnetization density – a product of the correlation function and Fourier transform of the magnetization density of the degrees of freedom. Using cold neutrons, which restricts the access to low scattering vectors only, we fail to provide evidence for a diffuse signal in energy-integrated experiments that would indicate correlations of magnetic dipoles (see Extended data). Instead, a signal of diffuse neutron scattering appears at higher scattering vectors accessible using thermal neutrons (Fig. 2**a-c**). The intensity distribution is zero at low scattering vectors $Q < 4$ Å$^{-1}$ but grows at higher scattering vectors, where a maximum is reached around $Q \sim 8$ Å$^{-1}$. This behaviour is precisely that expected from the Fourier transform of magnetic octupoles calculated from the type of ground-state wavefunction in Ce$_2$Sn$_2$O$_7$ (Extended data). The octupolar scattering increases in amplitude upon cooling below 1 K, concomitantly with the drop in the effective magnetic moment from $\sim 1.2\ \mu_B$ in the range between $T = 1$ and 10 K down to $\sim 0.7\ \mu_B$ at $\sim 0.1$ K (Fig. 3**a**). The temperature dependence of both these quantities can be understood as the evolution of the mixing of the wavefunctions of the two states of the single-ion doublet under the effect of correlations. The interactions mix the otherwise degenerate $|m_{J_z} = \pm 3/2\rangle$ states, forming new split eigenstates that minimize the energy of the system due to different magnetic dipole and octupole moment sizes.

For dominant octupole–octupole couplings, the octupole moment strengthens at the expense of the dipole moment, as shown in Fig. 1**e**, where the magnetic charge density of the ground-state doublet wavefunction is represented for different values of the dipole moment $J_z$.

In the following, we rationalize our observations of octupolar scattering on the basis of the Hamiltonian for rare-earth pyrochlores with 'dipole–octupole' doublets $\mathcal{H}_{DO} = \sum_{<ij>}[\mathcal{J}^{xx}\tau_i^x\tau_j^x + \mathcal{J}^{yy}\tau_i^y\tau_j^y + \mathcal{J}^{zz}\tau_i^z\tau_j^z + \mathcal{J}^{xz}(\tau_i^x\tau_j^z + \tau_i^z\tau_j^x)]$ (refs. 7-8). The doublet is modelled by pseudo-spin $S = 1/2$ operators $\vec{\tau}_i = (\tau_i^x, \tau_i^y, \tau_i^z)$, where the components $\tau_i^x$ and $\tau_i^z$ transform like magnetic dipoles while $\tau_i^y$ behaves as an octupole moment[7,8,24]. We use a simplified version of the generic Hamiltonian considering $\mathcal{J}^{xx} = \mathcal{J}^{xz} = 0$, which still captures the essential physics of octupolar phases. Within this approximation and using mean-field calculations, two sets of $\mathcal{J}^{yy}$ and $\mathcal{J}^{zz}$ parameters were found to reproduce the bulk magnetic properties at low temperature (see Fig. 3**a-b**, and the details of the calculations described in the Extended data). The drop of the effective magnetic moment below 1 K is well reproduced using a dominant octupole–octupole interaction of either $\mathcal{J}^{yy} = +\ 0.48\ \pm\ 0.06$ K or $\mathcal{J}^{yy} = -0.16 \pm 0.02$ K, and a small but finite dipole–dipole coupling $\mathcal{J}^{zz} = +0.03 \pm 0.01$ K. Positive and negative $\mathcal{J}^{yy}$ values correspond to phases where the magnetic charge density of the octupoles on a tetrahedron is constrained by '2-plus-2-minus' (see Fig. 2**d**) and 'all-plus-all-minus' rules, respectively, where 'plus' and 'minus' designate the two possible local mean-values of the octupolar operator associated with $\tau_i^y$. A dominant and negative $\mathcal{J}^{yy}$ implies that the octupoles are not frustrated and form a 'hidden' ordered ground state, which can be excluded based on the absence of signatures of a phase transition to a long-range ordered state in both our heat-capacity and thermal-neutron scattering measurements. Meanwhile, the case governed by $\mathcal{J}^{yy} > 0$ is frustrated and leads to an extensively degenerate manifold of *octupole ice* configurations (Fig. 2**d**), for which our Monte Carlo simulations predict diffuse neutron scattering at high scattering vectors shown in Fig. 2**e** and clearly distinguishable from that expected for a manifold of dipole ice configurations in $Ce_2Sn_2O_7$ (Fig. 2**f,g**). The calculated octupolar diffuse scattering has the high degree of structure in reciprocal space typical of ice phases, and its powder averaging reproduces the experimental data as shown in Fig. 2**c**. The *octupole ice* scattering is about several hundred times weaker than the value expected for spin-ice scattering, even with the small dipole-moment of $Ce_2Sn_2O_7$, which itself is much weaker than that of classical spin ices like $Ho_2Ti_2O_7$.

The finite value of $\mathcal{J}^{zz}$ obtained from the fit of the bulk magnetic properties (Fig. 3**a-b**) implies $\mathcal{J}^{\pm}/\mathcal{J}^{yy} = -\mathcal{J}^{zz}/4\mathcal{J}^{yy} \approx -0.015$, where $\mathcal{J}^{yy} = 0.48$ K and $\mathcal{J}^{\pm}$ follows the convention of the nearest-neighbour pseudo-spin 1/2 quantum spin ice model[29-30] after appropriate permutation of the pseudo-spin components[7-8]. The octupolar quantum spin ice is predicted to be stable in the range $0 < \mathcal{J}^{\pm}/\mathcal{J}^{yy} < 0.19$ (ref. 8), and it is also established that the U(1) quantum spin liquid is robust against perturbations and particularly stable in the regime where transverse exchange is frustrated ($\mathcal{J}^{\pm} < 0$) (ref. 30-34). Therefore, it is natural to conjecture that $\mathcal{J}^{\pm}/\mathcal{J}^{yy} \approx -0.015$ also falls in the octupolar quantum spin ice regime where, under applied magnetic field, theory predicts an Anderson–Higgs transition towards an ordered phase of dipole moments[8]. The critical field of this transition is of the order of 1.5 to 3 $h/\mathcal{J}^{yy}$, depending on the direction of the applied magnetic field $h$, which translates into values of 0.5 to 1 Tesla for $\mathcal{J}^{yy} = +0.48$ K. This is in good agreement with the appearance of magnetic Bragg peaks at similar field values in the elastic scattering of cold neutrons at 0.05 K (see Extended data). The same set of $\mathcal{J}^{zz}$ and $\mathcal{J}^{yy}$ parameters also reproduces the evolution of the heat capacity at low temperature under applied magnetic field (Fig. 3**c**).

We expect the emergent excitations in Ce$_2$Sn$_2$O$_7$ to be measurable using neutron spectroscopy, because $\mathcal{J}^{zz}$ allows neutron-active transitions between the two states of the doublet split by $\mathcal{J}^{yy}$. Low-energy neutron spectroscopy data indeed reveal the presence of low-energy excitations that are dipolar in nature (Fig. 4). In addition, the spectra measured in the correlated regime of Ce$_2$Sn$_2$O$_7$ do *not* display a resolution-limited excitation centred on a unique energy transfer – as would be expected for the 'hidden' ordered phase set by $\mathcal{J}^{yy} < 0$, but instead have the continuous character of fractionalized excitations in a quantum spin liquid (Fig. 4**c**,**f**). We attribute the continuum in Ce$_2$Sn$_2$O$_7$ to the gapped spinons of the U(1) quantum spin liquid, which are foreseen in inelastic neutron scattering of octupolar quantum spin ices[8], including when $\mathcal{J}^{\pm}$ is frustrated[34]. The continuum extends up to at least 0.3 meV (Fig. 4**e**); that is approximately one order of magnitude larger than the dominant exchange interaction $\mathcal{J}^{yy} \approx 0.05$ meV around which the spectrum is peaked, and both of these characteristics are consistent with theoretical predictions[4,35-37]. Moreover, the low-energy edge of the spectra shown in Fig. 4**b**,**e** is reminiscent of the threshold proposed for the effect of the emergent photon – the gapless excitation in the lattice gauge theory of quantum spin ice[2-4] – on the production of spinons, which leads to an abrupt onset in the density of states[37]. We note that the gapless photon itself is not expected to contribute to the inelastic neutron scattering of an octupolar quantum spin ice[8], at least not in the form of dipolar magnetic neutron scattering, and that the estimated photon bandwidth is of the order of the ring exchange $g \equiv 12\mathcal{J}^{\pm 3}/\mathcal{J}^{yy2} \approx 10^{-4}$ meV, which is much smaller than the energy scale of the inelastic scattering in Fig. 4.

The octupolar quantum spin ice scenario might also explain recent observations of quantum spin liquid dynamics in $Ce_2Zr_2O_7$ (ref. 38-39). In particular, the continuum of dipolar neutron scattering observed in ref. 39 indeed shows similar finite-energy centres and widths as in Fig. 4, and therefore can be used together with other measured properties that are similar in both materials to conclude that octupolar quantum spin ice is common among $Ce^{3+}$ pyrochlores. In their work[39], Gao *et al.* argue that $Ce_2Zr_2O_7$ should be in the π-flux phase of quantum spin ice[34], which is consistent with our estimates of exchange parameters ($\mathcal{J}^{\pm} < 0$), while our work further shows that the spin ice manifold is constructed from octupoles. In contrast, the interpretation of Gaudet *et al.*[38] that the inelastic neutron scattering in $Ce_2Zr_2O_7$ should be compared with the equal-time structure factor of a quantum spin ice, where the calculation[3] includes photon scattering but no spinons, appears inconsistent with our data and interpretation. However, the comparison of the two materials should be contrasted by the fact that the oxygen stoichiometry in $Ce_2Sn_2O_7$ is well controlled and precisely measured (see Extended data), so that the amount of non-magnetic $Ce^{4+}$ is known to be less than 1%, while the lattice constant of the studied samples of $Ce_2Zr_2O_7$ suggests about 10% of $Ce^{4+}$ (ref. 38). Such a level of site dilution is likely to have a significant effect on the correlations, even though the local moment of the ~ 90% of $Ce^{3+}$ is protected against disorder due to their Kramers' nature.

The discovery of the *octupole ice* in $Ce_2Sn_2O_7$ is remarkable for several reasons: it is an example of frustrated multipoles, showing convincing consistency with the expectations of a well-established 3-dimensional model – the U(1) quantum spin liquid; it is one of very few materials to enter phases primarily driven by higher multipolar interactions[5-6], and certainly a unique example of a quantum liquid in this context. Finally, degrees of freedom in the *octupole ice* are quantum objects whose dual 'dipole–octupole' nature can be controlled using magnetic fields[8] or – as our results demonstrate – as a function of temperature, which gives interesting perspectives in the field of quantum many-body systems and future related technologies.

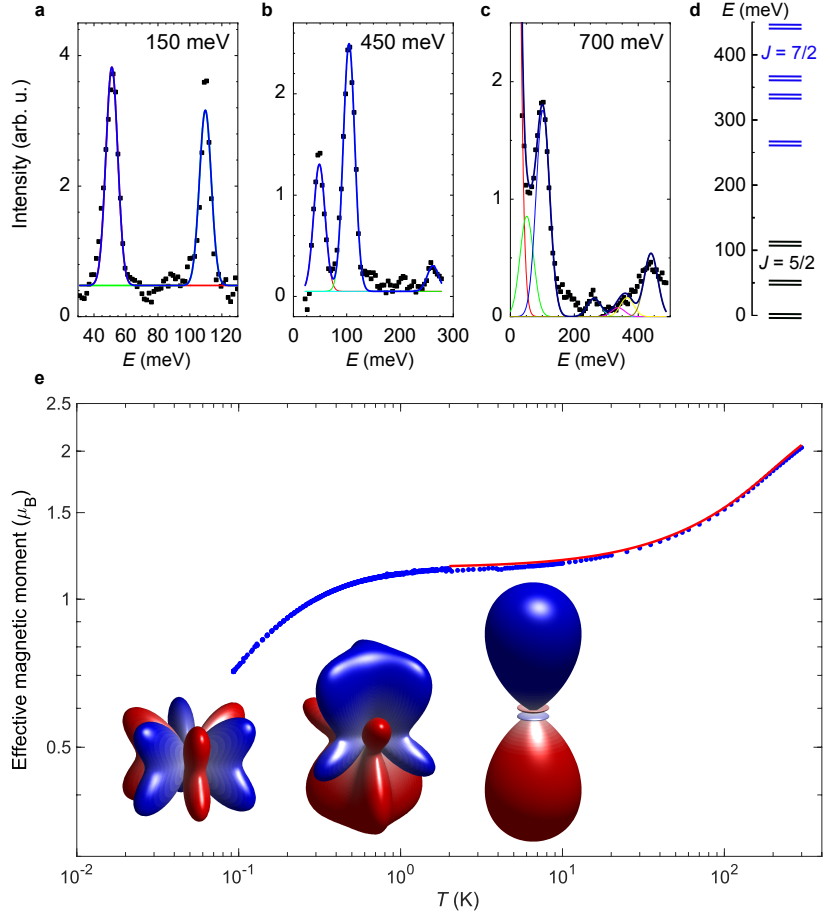

**Figure 1 | Dipole–octupole degrees of freedom.** Energy spectra measured at 6 K (**a-c**) and recorded using incident neutrons of different energies as specified on the top right-hand corner of each panel. The solid lines are fits of the experimental data (black squares) using Gaussian peaks (peaks within each spectrum are constrained to the same width), and the resulting energy positions are shown in panel **d**. The energies and intensities of the magnetic excitations were used to adjust the six parameters of a crystal field Hamiltonian, which gives the ground-state doublet wavefunction $|\pm\rangle = 0.87|^2F_{5/2},\pm 3/2\rangle \pm 0.46|^2F_{5/2},\mp 3/2\rangle \mp 0.15|^2F_{7/2},\pm 3/2\rangle - 0.01|^2F_{7/2},\mp 3/2\rangle$. We note that although both $J = 5/2$ and $J = 7/2$ multiplets contribute to this wavefunction, the composition is dominated by the ground multiplet $J = 5/2$ (see Extended data), as is the case for all the doublets drawn in black on panel **d**, while those represented in blue are dominated by the higher multiplet $J = 7/2$. Panel **e** shows the temperature dependence of the effective magnetic moment $\mu_{eff} \propto (\chi \times T)^{1/2}$ determined experimentally from measurements of the magnetization $M$ performed under a weak magnetic field $H$, which gives the magnetic susceptibility $\chi = M/H$ in the linear-field regime. The solid red line is the effective magnetic moment calculated from the parameters fitted from the neutron data shown in panels **a-c**. The inserts in panel **e** represent the magnetic charge density calculated from the type of ground state wavefunction of $Ce^{3+}$ determined from the fit of the neutron data. In the temperature range between ~1 K and ~10 K, this wavefunction results in a dipole moment $J_z = 3/2$ shown on the right, which would persist down to the lowest temperatures in the absence of correlations. However, under the effect of octupole–octupole correlations, this quantum object encoded in the local ground-state wavefunction has the potential to reveal an octupole moment at the expense of the dipole moment. This is shown in the middle and left drawings in panel **e**, where the dipole moment $J_z$ is half of that in the uncorrelated regime and zero, respectively.

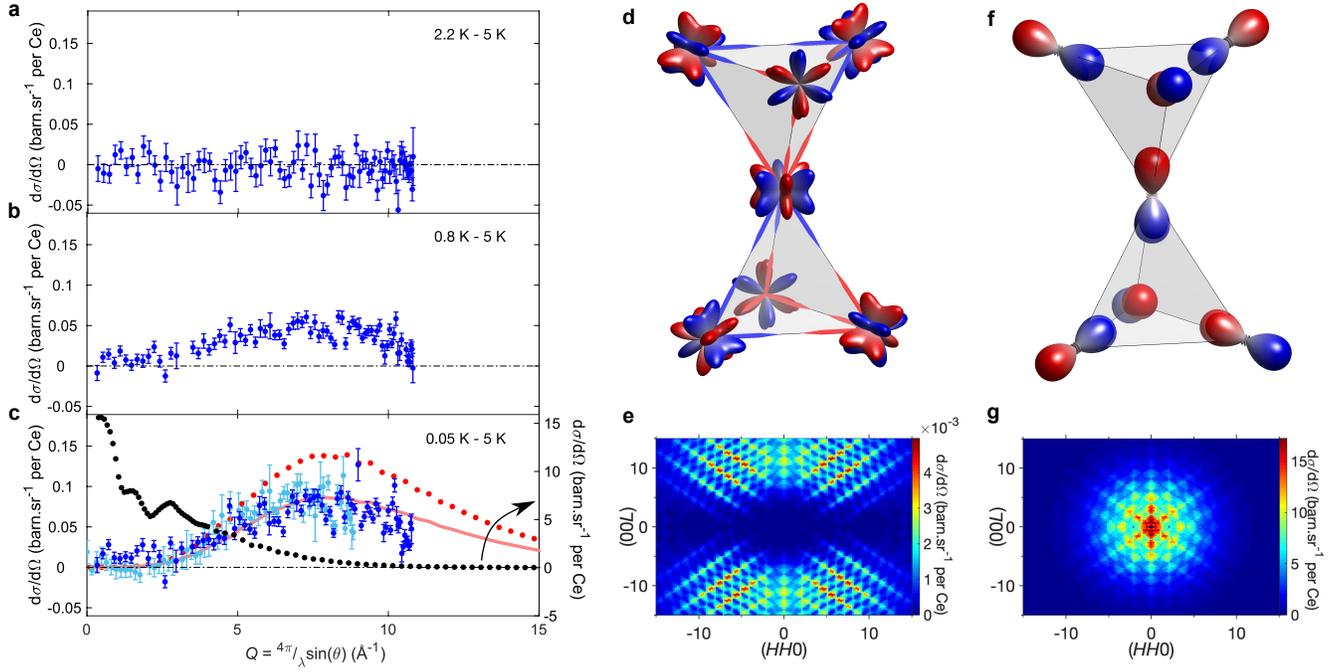

**Figure 2 | Octupole–octupole correlations.** In panel **a-c** we present the diffuse scattering (blue points with error bars corresponding to ±1 standard error) obtained from the difference between neutron diffraction patterns measured at 5 K and at a lower temperature indicated on each panel. Measurements were performed on two different diffractometers using thermal neutrons of wavelengths λ = 1.15 Å (dark blue points on panels **a-c**) or λ = 1.37 Å (light blue points on panel **c**), which give access to the large scattering vectors required to observe scattering by magnetic octupoles. Diffuse scattering is expected for short-range correlated phases, which on the pyrochlore lattice are ice phases. Given the dipole–octupole nature of the degrees of freedom shown in Fig. 1, ice phases in $Ce_2Sn_2O_7$ can be based on either octupoles or dipoles, whose local configurations are shown in panels **d** and **f**, respectively, where the blue and red colours encode regions of opposite magnetization. Magnetic dipoles respecting the '2-in-2-out' ice rule on each tetrahedron are shown in panel **f**. In panel **d**, the octupolar '2-plus-2-minus' ice rule is further highlighted using blue/red colours on the edges of the lattice along the corresponding magnetization directions of the closest octupole. We note that a variant of *octupole ice* configurations can be considered, where each octupole is rotated by 30 degrees around the local trigonal direction relative to what is shown here. In panels **e** and **g**, we show the diffuse scattering calculated in the (*HHL*) plane of reciprocal space using Monte Carlo simulations for the *octupole ice* and a spin ice of magnetic dipoles with first neighbour interactions only, respectively (see Extended data). Note that the spin ice pattern (panel **g**) is displayed over a much larger area of reciprocal space than usual, but the typical features can be discerned in the central region. The powder average of the diffuse scattering calculated for the *octupole ice* and spin ice configurations is shown respectively with red and black points in panel **c**, while the solid red line represents the same calculation for the octupole ice but scaled (× 0.625) onto the experimental data. The octupolar scattering measured at 0.05 K is about two thirds of the intensity of the zero-temperature calculation assuming a full octupolar moment, which is a remarkable agreement giving strong support to the existence of *octupole ice* correlations in this material. Note the different scales used to display the octupolar (left scale) and dipolar (right scale) scattering in $Ce_2Sn_2O_7$.

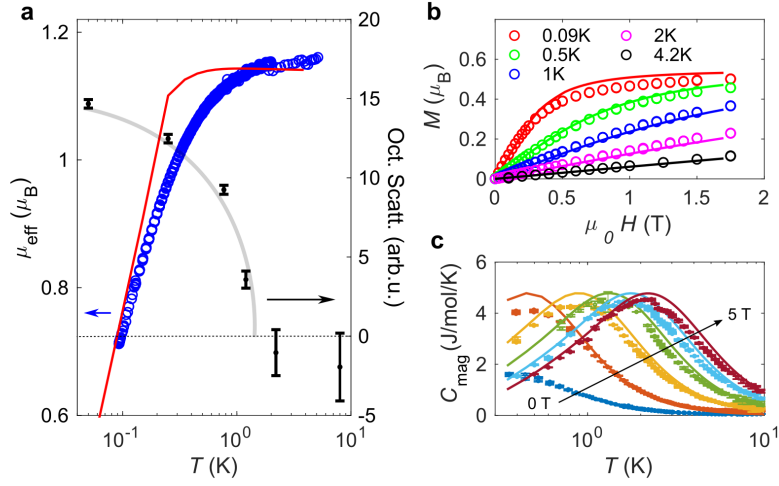

**Figure 3 | Ground-state properties.** Open circles in panels **a** and **b** show the effective magnetic moment $\mu_{\text{eff}} \propto (\chi \times T)^{1/2}$ as a function of temperature $T$ and the magnetization $M$ as a function of magnetic field $H$, respectively (ref. 28). The lines are the best fits using the Hamiltonian for rare-earth pyrochlores with 'dipole–octupole' doublets $\mathcal{H}_{DO} = \sum_{<ij>}[\mathcal{J}^{xx}\tau_i^x\tau_j^x + \mathcal{J}^{yy}\tau_i^y\tau_j^y + \mathcal{J}^{zz}\tau_i^z\tau_j^z + \mathcal{J}^{xz}(\tau_i^x\tau_j^z + \tau_i^z\tau_j^x)]$, where the components $\tau_i^x$ and $\tau_i^z$ of the pseudo-spin operator $\vec{\tau}_i = (\tau_i^x, \tau_i^y, \tau_i^z)$ transform like magnetic dipoles, while $\tau_i^y$ behaves as an octupole moment (refs. 7-8). We consider $\mathcal{J}^{xx} = \mathcal{J}^{xz} = 0$ and we show here the fits for the hypothesis $\mathcal{J}^{yy} > 0$, which gives the octupole–octupole interaction $\mathcal{J}^{yy} = +0.48 \pm 0.06$ K and the dipole–dipole interaction $\mathcal{J}^{zz} = +0.03 \pm 0.01$ K. The discrepancy between the experimental effective moment and the fitted curve is due to the mean-field nature of the model, which is sufficient to evaluate the main ingredients of the Hamiltonian but cannot account exactly for all details in the quantum regime, and the logarithmic temperature scale of panel **a** makes this particularly apparent. The same set of parameters also reproduces the field dependence of the magnetic contribution to the heat capacity, shown in panel **c** for field intervals of 1 T (points with error bars corresponding to ±1 standard error), as obtained from measurements of the heat capacity of $Ce_2Sn_2O_7$ and of the non-magnetic analogue $La_2Sn_2O_7$. The temperature dependence of the integrated octupolar scattering is presented in panel **a** (black points with error bars corresponding to ±1 standard error), showing the increase of octupolar scattering upon cooling in the correlated regime and at the expense of the dipole moment represented by the effective magnetic moment (the grey curve is a guideline). The dominant octupole–octupole interaction $\mathcal{J}^{yy}$ and the small finite value of the dipole–dipole coupling $\mathcal{J}^{zz}$ indicate that $Ce_2Sn_2O_7$ is in the octupolar quantum spin ice phase ($\mathcal{J}^{\pm}/\mathcal{J}^{yy} = -\mathcal{J}^{zz}/4\mathcal{J}^{yy} \approx -0.015$). In this regime, we anticipate the excitation spectrum to have the continuous character expected for the fractionalized excitations of a quantum spin liquid, which we show is the case in Figure 4.

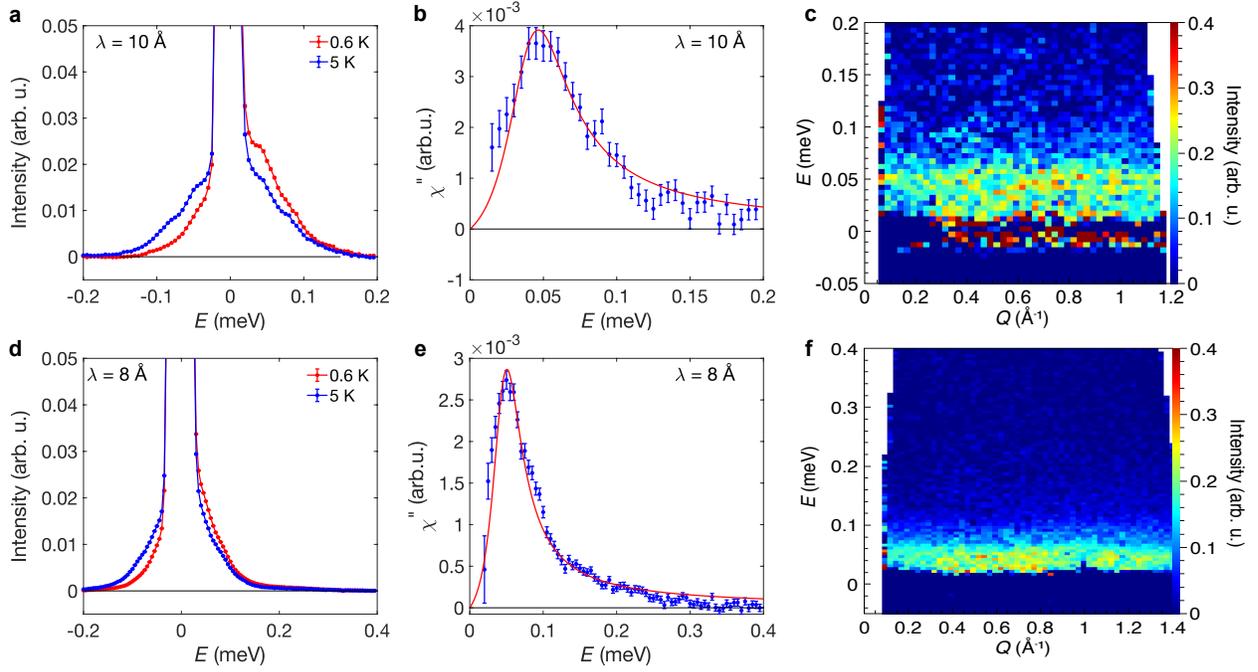

**Figure 4 | Magnetic excitations.** Left panels show low-energy inelastic neutron scattering data measured at 0.6 K (red) and 5 K (blue) using incident energies $E_i$ of 0.82 meV (**a**, $\lambda$ = 10 Å) and 1.28 meV (**d**, $\lambda$ = 8 Å), which give energy resolutions of 0.011 meV and 0.020 meV at the elastic line, respectively. These spectra as a function of energy transfer $E$ are integrated over the entire range of accessible scattering vectors $\mathbf{Q}$ shown on the ($E$,$\mathbf{Q}$) maps (see right panels **c** and **f** for $\lambda$ = 10 Å and $\lambda$ = 8 Å, respectively). Assuming that the difference scattering $S(E)$ between 0.6 K and 5 K is entirely magnetic, we show in the middle panels (**b** and **e** for $\lambda$ = 10 Å and $\lambda$ = 8 Å, respectively) the imaginary part of the dynamic spin susceptibility (blue points with error bars corresponding to ±1 standard error) calculated as $\chi''(E) = S(E) \times [1 - \exp(-E/k_B T)]$, where $E$ is the neutron energy transfer and $k_B$ is the Boltzmann's constant. We model the magnetic scattering using $\chi''(E) = \Gamma E/[(E-\Delta)^2 + \Gamma^2]$ (red line) – a Lorentzian peak shape where $\Gamma$ and $\Delta$ characterise the damping and energy centre, respectively, which we use as phenomenological parameters to compare with results on $Ce_2Zr_2O_7$ (ref. 39). The width and energy center determined for $\lambda$ = 10 Å are $\Gamma$ = 0.025±0.003 meV and $\Delta$ = 0.039±0.003 meV, respectively ($\Gamma$ = 0.024±0.001 meV and $\Delta$ = 0.045±0.002 meV for $\lambda$ = 8 Å). Panels **c** and **f** show difference ($E$,$\mathbf{Q}$) maps (0.6 − 5 K) summarizing the wavevector dependence of the spin excitations as a function of energy. These plots give strong support to the existence of a gapped continuum of fractionalized spin excitations, which we attribute to the spinon excitations of the octupolar quantum spin ice[8,34] (see main text). The phenomenological form employed to fit the spectra in panels **b** and **e** happens to capture the features expected from theory for the spinon excitations in a quantum spin ice (onset, peak, and extent)[4,35-37].

**References**

1. Savary, L & Balents, L. Quantum spin liquids: a review. *Rep. Prog. Phys.* **80**, 016502 (2016).

2. Hermele, M., Fisher, M. P. A. & Balents, L. Pyrochlore photons: The $U(1)$ spin liquid in a $S = ½$ three-dimensional frustrated magnet. *Phys. Rev. B* **69**, 064404 (2004).

3. Benton, O., Sikora, O. & Shannon, N. Seeing the light: Experimental signatures of emergent electromagnetism in a quantum spin ice. *Phys. Rev. B* **86**, 075154 (2012).

4. Gingras, M. J. P. & McClarty, P. A. Quantum spin ice: a search for gapless quantum spin liquids in pyrochlore magnets. *Rep. Prog. Phys.* **77**, 056501 (2014).

5. Kuramoto, Y., Kusunose, H. & Kiss, A. Multipole Orders and Fluctuations in Strongly Correlated Electron Systems. *J. Phys. Soc. Jpn.* **78**, 072001 (2009).

6. Santini, P. *et al.* Multipolar interactions in f-electron systems: The paradigm of actinide dioxides. *Rev. Mod. Phys.* **81**, 807–863 (2009).

7. Huang, Y.-P., Chen, G. & Hermele, M. Quantum Spin Ices and Topological Phases from Dipolar-Octupolar Doublets on the Pyrochlore Lattice. *Phys. Rev. Lett.* **112**, 167203 (2014).

8. Li, Y.-D. & Chen, G. Symmetry enriched U(1) topological orders for dipole-octupole doublets on a pyrochlore lattice. *Phys. Rev. B* **95**, 041106 (2017).

9. Ikeda, H. *et al.* Emergent rank-5 nematic order in $URu_2Si_2$. *Nature Phys.* **8**, 528–533 (2012).

10. Paddison, J. A. M. *et al.* Hidden order in spin liquid $Gd_3Ga_5O_{12}$. *Science* **350**, 179–181 (2015).

11. Lovesey, S. W., Khalyavin, D. D. and Staub, U. Ferro-type order of magneto-electric quadrupoles as an order-parameter for the pseudo-gap phase of a cuprate superconductor. *J. Phys. Condens. Matter* **27**, 292201 (2015).

12. Fechner, M., Fierz, M. J. A., Thöle, F., Staub, U. and Spaldin, N. A. Quasistatic magnetoelectric multipoles as order parameter for pseudogap phase in cuprate superconductors. *Phys. Rev. B* **93**, 174419 (2016).

13. Zhao, L. *et al.* Evidence of an odd-parity hidden order in a spin–orbit coupled correlated iridate. *Nature Phys.* **12**, 32–36 (2016).

14. Watanabe, H. and Yanase, Y. Magnetic hexadecapole order and magnetopiezoelectric metal state in $Ba_{1-x}K_xMn_2As_2$. *Phys. Rev. B* **96**, 064432 (2017).

15. Hayami, S., Yanagi, Y., Kusunose, H. and Motome, Y. Electric Toroidal Quadrupoles in the Spin-Orbit-Coupled Metal $Cd_2Re_2O_7$. *Phys. Rev. Lett.* **122**, 147602 (2019).

16. Spaldin, N. A., Fiebig, M. and Mostovoy, M. The toroidal moment in condensed-matter physics and its relation to the magnetoelectric effect. *J. Phys. Condens. Matter* **20**, 434203 (2008).

17. Di Matteo, S. and Norman, M. R. Orbital currents, anapoles, and magnetic quadrupoles in CuO. *Phys. Rev. B* **85**, 235143 (2012).

18. Wen, X-G. Quantum orders and symmetric spin liquids. *Phys. Rev. B* **65** 165113 (2002).

19. Knolle, J. & Moessner, R. A Field Guide to Spin Liquids. *Annu. Rev. Condens. Matter Phys.* **10**, 451–472 (2019).

20. Han, T.-H. *et al.* Fractionalized excitations in the spin-liquid state of a kagome-lattice antiferromagnet. *Nature* **492**, 406–410 (2012).

21. Plumb, K. W. *et al.* Continuum of quantum fluctuations in a three-dimensional S = 1 Heisenberg magnet. *Nature Phys.* **15**, 54–59 (2019).

22. Kitagawa, K. *et al.* A spin-orbital-entangled quantum liquid on a honeycomb lattice. *Nature* **554**, 341–345 (2018).

23. Paddison, J. A. M. *et al.* Continuous excitations of the triangular-lattice quantum spin liquid YbMgGaO$_4$. *Nature Phys.* **13**, 117–122 (2017).

24. Rau, J. G. & Gingras, M. J. P. Frustrated Quantum Rare-Earth Pyrochlores. *Annu. Rev. Condens. Matter Phys.* **10**, 357–386 (2019).

25. Castelnovo, C., Moessner, R. & Sondhi, S. L. Spin Ice, Fractionalization, and Topological Order. *Annu. Rev. Condens. Matter Phys.* **3**, 35–55 (2012).

26. Rau, J. G. & Gingras, M. J. P. Magnitude of quantum effects in classical spin ices. *Phys. Rev. B* **92**, 144417 (2015).

27. Onoda, S. & Tanaka, Y. Quantum Melting of Spin Ice: Emergent Cooperative Quadrupole and Chirality. *Phys. Rev. Lett.* **105**, 047201 (2010).

28. Sibille, R. *et al.* Candidate Quantum Spin Liquid in the Ce$^{3+}$ Pyrochlore Stannate Ce$_2$Sn$_2$O$_7$. *Phys. Rev. Lett.* **115**, 097202 (2015).

29. Curnoe, S. H. Structural distortion and the spin liquid state in Tb$_2$Ti$_2$O$_7$. *Phys. Rev. B* **78**, 094418 (2008).

30. Lee, S., Onoda, S. and Balents, L. Generic quantum spin ice. *Phys. Rev. B* **86**, 104412 (2012).

31. Benton, O., Jaubert, L. D. C., Singh, R. R. P., Oitmaa, J. and Shannon, N. Quantum Spin Ice with Frustrated Transverse Exchange: From a π-Flux Phase to a Nematic Quantum Spin Liquid. *Phys. Rev. Lett.* **121**, 067201 (2018).

32. Chen, G. Spectral periodicity of the spinon continuum in quantum spin ice. *Phys. Rev. B* **96**, 085136 (2017).

33. Taillefumier, M., Benton, O., Yan, H., Jaubert, L. D. C. and Shannon, N. Competing Spin Liquids and Hidden Spin-Nematic Order in Spin Ice with Frustrated Transverse Exchange. *Phys. Rev. X* **7**, 041057 (2017).

34. Li, Y.-D. and Chen, G. Non-spin-ice pyrochlore U(1) quantum spin liquid: Manifesting mixed symmetry enrichments. *arXiv:1902.07075* (2019).

35. Huang, C.-J., Deng, Y., Wan Y. and Meng Z.-Y. Dynamics of topological excitations in a model quantum spin ice. *Phys. Rev. Lett.* **120**, 167202 (2018).

36. Udagawa, M. and Moessner, R. Spectrum of itinerant fractional excitations in quantum spin ice. *Phys. Rev. Lett.* **122**, 117201 (2019).

37. Morampudi, S. D., Wilczek, F. and Laumann, C. R. Spectroscopy of spinons in Coulomb quantum spin liquids. *arXiv:1906.01628* (2019).

38. Gaudet, J. *et al.* Quantum spin ice dynamics in the dipole-octupole pyrochlore magnet Ce$_2$Zr$_2$O$_7$. *Phys. Rev. Lett.* **122**, 187201 (2019).

39. Gao, B. *et al.* Experimental signatures of a three-dimensional quantum spin liquid in effective spin-1/2 Ce$_2$Zr$_2$O$_7$ pyrochlore. *Nature Phys.* **15**, 1052–1057 (2019).

**Acknowledgements**

This work is based on experiments performed at the Swiss spallation neutron source SINQ, Paul Scherrer Institute, Villigen, Switzerland. Experiments at the ISIS Neutron and Muon Source were supported by a beamtime allocation RB1510524 from the Science and Technology Facilities Council. Additional neutron scattering experiments were also carried out at the Institut Laue Langevin, ILL, Grenoble, France. We thank M. Kenzelmann, O. Zaharko, Y.-P. Huang, M. Müller and C. Mudry for useful discussions. We thank C. Paulsen for the use of his magnetometers; P. Lachkar for help with the PPMS; B. Fåk for assisting on IN4c, E. Pomjakushina and D. Gawryluk for providing help and access to the solid-state chemistry laboratory; and M. Zolliker and M. Bartkowiak for dedicated work on running the dilution refrigerators at SINQ. R.S. thanks N. Shannon for fruitful exchanges and great hospitality at OIST; as well as H. Yan for enlightening conversations. We acknowledge funding from the Swiss National Science Foundation (Grant No. 200021_179150 and Fellowship No. P2EZP2 178542).

**Author contributions**

Project and experiments were designed by R.S. Sample preparation and characterization were performed by R.S. and V.Porée. Neutron scattering experiments were carried out by R.S., E.L., V.Porée and T.F. with V.Pomjakushin, R.E., T.P., J.O., A.W., C.R., T.H., D.A.K., G.N. and L.K. as local contacts. Measurements of bulk properties were performed by E.L. Experimental data were analysed by R.S., N.G., E.L., V.Porée, V.Pomjakushin, S.P. and T.F. Calculations were performed by N.G. and S.P. Graphical representations of the magnetic charge densities shown in Fig. 1**f-h** and 2**d**,**f** were produced by N.G., while all the other elements of the main figures were produced by R.S. and V.Porée based on (i) neutron scattering data taken by R.S. and T.F. (Fig. 1**a-c** and 3**d**), R.S. and V.Porée (Fig. 2**a-c** and 3**b**) and V.Porée and E.L. (Fig. 4), (ii) macroscopic measurements performed by E.L. (Fig. 1**e** and 3**a-c**), and (iii) calculations performed by S.P. (Fig. 2**c**,**e**,**g**). The paper was written by R.S. with feedback from all authors, especially N.G., E.L., S.P. and T.F.

**Competing financial interests**

The authors declare no competing financial interests.

**Methods**

**Sample preparation**

A 28-grams polycrystalline sample of $Ce_2Sn_2O_7$ was prepared by the solid-state reaction described in Ref. 31. $CeO_2$ was added to appropriate amounts of $SnO_2$ and Sn (the $Sn^{4+}/Sn^0$ ratio is imposed in order to reduce all the $Ce^{4+}$ cations to $Ce^{3+}$). The resulting mixture was carefully mixed in a rotary blender for several hours under air, in order to prevent agglomeration of metallic tin. The mixture was then placed into an alumina crucible for several heat treatments of 24-hours each until complete reaction. The first heat treatment was at 900 Celsius and the following ones at 1000 Celsius, all under flowing argon after long purging of the tube furnace, and using slow heating and cooling ramps (1 Celsius per minute).

As a non-magnetic reference material, La$_2$Sn$_2$O$_7$ was synthesised from La$_2$O$_3$ calcinated at 800 Celsius for several hours and mixed with SnO$_2$ in a rotary blender. The phase-pure lanthanum stannate was obtained by heating under air at 1200 Celsius for 72 hours, with intermediate regrinding each 24 hours.

Our sample of Ce$_2$Sn$_2$O$_7$ was characterized by high-resolution neutron powder diffraction, pair distribution function, and thermogravimetric measurements, all indicating the presence of a single phase of formula Ce$_2$Sn$_2$O$_{7.00\pm0.01}$ in our sample (see Extended data), thus forming a perfect pyrochlore lattice of Ce$^{3+}$ ions.

**Bulk properties measurements**

Magnetization data were measured in the temperature range from 1.8 to 300 K in an applied magnetic field of 0.01 Tesla using a Quantum Design MPMS-XL SQUID magnetometer. Additional magnetization, and ac-susceptibility (see Extended data), measurements were made as a function of temperature and field, from $T$ = 0.07 to 4.2 K and from $\mu_0 H$ = 0 to 8 T, using SQUID magnetometers equipped with a miniature dilution refrigerator developed at the Institut Néel, CNRS in Grenoble. The heat capacity of pelletized samples of both Ce$_2$Sn$_2$O$_7$ and La$_2$Sn$_2$O$_7$ was measured down to 0.3 K using a Quantum Design PPMS.

**Neutron scattering experiments**

The inelastic high-energy neutron scattering experiment was performed on the MAPS spectrometer[32] at the ISIS Neutron Facility in Didcot, UK. Data were evaluated using HORACE software suite[33]. Each of the Ce$_2$Sn$_2$O$_7$ and La$_2$Sn$_2$O$_7$ samples were mounted inside aluminium foil pouches, which were then mounted in annular geometry inside aluminium cans. The crystal field splitting of Ce$^{3+}$ was evaluated by comparing Ce$_2$Sn$_2$O$_7$ and La$_2$Sn$_2$O$_7$ data, which allows to distinguish among contributions from crystal field and phonon excitations. The spectra of magnetic neutron scattering intensity as a function of energy transfer presented in Fig. 1**a-c** were obtained from low-angle data from Ce$_2$Sn$_2$O$_7$ with high-angle data from Ce$_2$Sn$_2$O$_7$ subtracted, using a scale factor determined from La$_2$Sn$_2$O$_7$ data (Ref. 34). Unlabelled signals in the spectra of Fig. 1**a-c** are attributed to residual phonon scattering from Ce$_2$Sn$_2$O$_7$ or hydrogen-containing contaminants formed through exposure to air. Note that the range of scattering vectors $Q$ is not constant as a function of energy transfer in the spectra of Fig. 1**a-c**, thus peak intensities have to be extrapolated to $Q$ = 0 by correction for the magnetic form factor. The SPECTRE program[35] was used to fit the neutron data to the crystal-field Hamiltonian.

The energy-integrated thermal-neutron scattering experiment was performed on the HRPT diffractometer at SINQ, PSI in Villigen, Switzerland[36]. Measurements were performed with the sample enclosed in a copper can and cooled down using a dilution refrigerator. Each pattern was measured during 6 to 12 hours, and the differences were made by subtracting data recorded at 5 K with the exact same statistics, resulting in the experimental data shown in Fig. 2**a-c**. A similar experiment was repeated on the D20 diffractometer at the Institut Laue Langevin in Grenoble, France[37], resulting in the second set of data points shown in Fig.2**c**.

The inelastic low-energy neutron scattering experiment was performed on the IN5 spectrometer at Institut Laue Langevin in Grenoble, France[38]. Measurements were performed with the sample enclosed in a copper can and cooled down using a dilution refrigerator.

## Calculations

For the visualisation of the magnetic degrees of freedom (Fig. 1**e** and Fig. 2**d**,**f**), we represent the magnetization by its corresponding magnetic charge distribution defined as $\rho_{magnetic}(\boldsymbol{r}) = -\nabla \cdot m(\boldsymbol{r})$. This charge distribution is calculated using the spherical Racah tensor operators on the wavefunction of the ground state[39], projected on the ground multiplet $^2F_{5/2}$.

Fits of the bulk magnetic properties are based on calculations carried out on the basis of a mean-field treatment of a Hamiltonian, considering the dipolar exchange as well as an octupolar coupling between the crystal-electric field ground doublet states of the $Ce^{3+}$ ion. This Hamiltonian is written in terms of a pseudospin 1/2 spanning these states.

The elastic neutron cross sections of the octupole ice (Fig. 2**e**) and dipole ice (Fig. 2**g**) for $Ce_2Sn_2O_7$ were calculated numerically using Monte Carlo simulations performed assuming 100 random realisations of the corresponding ice lattices (see Extended data). Each random lattice contained 5488 sites.

## Data availability

The data that support the plots within this paper and other findings of this study are available from the corresponding authors upon reasonable request (R.S. for all of the experimental data, and S.P. for the mean-field fits of the bulk properties and Monte Carlo simulations). The datasets for the inelastic neutron scattering experiment on MAPS and MERLIN are available from the ISIS Neutron and Muon Source Data Catalogue[40-41]. The datasets for the polarized neutron scattering experiment on D7 (see extended data), the thermal-neutron inelastic measurements on IN4 (see extended data) and the diffraction experiment on D20 are available from the Institute Laue-Langevin data portal[42-45].

## References (methods)

31. Tolla, B. *et al.* Oxygen exchange properties in the new pyrochlore solid solution $Ce_2Sn_2O_7$ – $Ce_2Sn_2O_8$, *Comptes rendus de l'académie des sciences. Serie IIc, chimie* **2**, 139–146 (1999).

32. Ewings, R. A. *et al.* Horace: Upgrade to the MAPS neutron time-of-flight chopper spectrometer. *Rev. Sci. Inst.* **90**, 035110 (2019).

33. Ewings, R. A. *et al.* Horace: Software for the analysis of data from single crystal spectroscopy experiments at time-of-flight neutron instruments. *Nucl. Instrum. Methods Phys. Res. Sect. A* **834**, 132–142 (2016).

34. Princep, A. J., Prabhakaran, D., Boothroyd, A. T. and Adroja, D. T. Crystal-field states of $Pr^{3+}$ in the candidate quantum spin ice $Pr_2Sn_2O_7$. *Phys. Rev. B* **88**, 104421 (2013).

35. Boothroyd, A. T. SPECTRE, a program for calculating spectroscopic properties of rare earth ions in crystals, (1990-2018).

36. Fischer, P. *et al.* High-resolution powder diffractometer HRPT for thermal neutrons at SINQ. *Physica B* **146**, 276–278 (2000).

37.  Hansen, T. C., Henry, P. H., Fischer, H. E., Torregrossa, J. and Convert, P. The D20 instrument at the ILL: a versatile high-intensity two-axis neutron diffractometer. Measurement Science and Technology **19**, 034001 (2008).

38.  Ollivier, J. & Mutka, H. IN5 cold neutron time-of-flight spectrometer, prepared to tackle single crystal spectroscopy. *J. Phys. Soc. Jpn* **80**, SB003 (2011).

39.  Kusunose, H. Description of Multipole in f-Electron Systems. *J. Phys. Soc. Jpn* **77**, 64710 (2008).

40.  SIBILLE Romain *et al.* (2015). Crystal electric field excitations of pyrochlore quantum antiferromagnet $Ce_2Sn_2O_7$, STFC ISIS Neutron and Muon Source, https://doi.org/10.5286/ISIS.E.RB1510524

41.  SIBILLE Romain *et al.* (2019). Hidden correlations in $Ce_2Sn_2O_7$, STFC ISIS Neutron and Muon Source, https://doi.org/10.5286/ISIS.E.RB1910467

42.  SIBILLE Romain *et al.* (2015). Investigation of the exotic low temperature groundstate of the $S_{eff}$ = 1/2 pyrochlore antiferromagnet $Ce_2Sn_2O_7$. Institut Laue-Langevin (ILL) https://doi.org/10.5291/ILL-DATA.5-32-809

43.  SIBILLE Romain *et al.* (2016). Investigation of the exotic low temperature groundstate of the $S_{eff}$ = 1/2 pyrochlore antiferromagnet $Ce_2Sn_2O_7$ – continuation. Institut Laue-Langevin (ILL) https://doi.org/10.5291/ILL-DATA.5-32-834

44.  SIBILLE Romain *et al.* (2015). Crystal electronic field excitations of pyrochlore antiferromagnet $Ce_2Sn_2O_7$. Institut Laue-Langevin (ILL) https://doi.org/10.5291/ILL-DATA.4-03-1716

45.  SIBILLE Romain *et al.* (2019). Hidden Correlations in a Quantum Liquid of Magnetic Octupoles. Institut Laue-Langevin (ILL) https://doi.org/10.5291/ILL-DATA.DIR-166